\documentclass{ws-ijgmmp}
\usepackage{cite}
\usepackage{graphicx}
\usepackage{graphics}
\usepackage{float}
\usepackage{amsmath}
\usepackage{placeins}
\usepackage{enumitem}
\usepackage[usenames]{color} 
\usepackage[linktocpage,colorlinks]{hyperref}
\hypersetup{colorlinks=true, citecolor=darkblue, linkcolor=darkblue, 
	urlcolor = darkblue}
\definecolor{darkblue}{rgb}{0,0,0.5}
\newcommand{\ber}{\begin{eqnarray}}
	\newcommand{\eer}{\end{eqnarray}}
\begin{document}
	\markboth{S. Ali, S. Khan, S. Sattar, \& A. Abebe}{The R\'{e}nyi holographic dark energy model in Chern-Simons gravity: some cosmological implications}
	
	\catchline{}{}{}{}{}

	\title{The R\'{e}nyi holographic dark energy model in Chern-Simons gravity: some cosmological implications}
	\author{Sarfraz Ali\footnote{sarfaraz.ali@ue.edu.pk}}
	\address{Department of Mathematics, University of Education, Lahore, Pakistan}
	\author{Sarfaraz Khan\footnote{sfz4ukhan@gmail.com}}
	\address{Riphah International University Faisalabad Campus, Pakistan}
	\author{Sadia Sattar\footnote{sadiasattar30@yahoo.com}}
	\address{Department of Mathematics, University of Sargodha, Sargodha, Pakistan}
	\author{Amare Abebe\footnote{amare.abbebe@gmail.com}}
	\address{Center for Space Research, North-West University, Mahikeng 2745, South Africa\\
		National Institute for Theoretical and Computational Sciences (NITheCS), South Africa}
	
	\maketitle
	\begin{history}
		\received{(Day Month Year)}
		\revised{(Day Month Year)}
	\end{history}
	
	\begin{abstract}
		In this paper, we study the R\'{e}nyi holographic dark energy model in the context of the Chern-Simons  model of modified gravity theory. Different cosmological parameters such as energy density, deceleration parameter, equation of state, square of sound speed and cosmological plane are discussed using the Friedmann-Lema\^{i}tre-Robertson-Walker spacetime background. Two separate solutions of CS field equations arise. The R\'{e}nyi HDE model shows the transition from deceleration to acceleration phase which is fully consistent with the observational data while the second case represents a decelerated phase of expansion only. Our results, based on the results of the equation of state obtained, predict that the universe is under the influence of dark energy and therefore in an accelerated expansion phase. On the other hand, the second-case solution shows the influence of $\Lambda$CDM. In both cases, $\omega <0,~ \omega^{\prime}<0$ indicated that the R\'{e}nyi HDE model is in the freezing region and cosmic expansion is more accelerating in the context of CS modified gravity. It is also observed that the value of the squared sound speed, $c^2_s$, is positive $\forall z\in \Re $, a sufficient condition for the stability of the system. Hence it is concluded that the R\'{e}nyi HDE model is supported by the results of general relativity in the framework of CS modified gravity.
	\end{abstract}
	\ccode{PACS: 04.20.-q, 04.50.Kd,99.80-k, 98.80-Jk.}
	\section{Introduction}
	
	The astonishing revelation of the accelerated expansion of the universe is one of the energizing advances in cosmology over most recent decades \cite{[a],[a1],[a2]}. It has been demonstrated by perceptions that the universe is flat, profoundly homogeneous and isotropic on large scales. Then again, dark energy is an extraordinary type of energy with negative pressure that has been recommended to describe the cosmic acceleration. On behalf of the standard model of cosmology, dark energy is about $68\%$ of the total content of the universe and contribution of mass energies in the form of dark matter and ordinary matters are about $27\%$ and $5\%$ respectively.
	
	The Chern-Simons (CS) modified gravity is a notably powerful augmentation of general relativity (GR) that considers non-minimal coupling between a gravitational scalar degree of freedom and the topological Pontryagin density in four dimensions \cite{[b]}. This theory arises due to the anomaly cancellation in curved spacetimes, string theory, and particle physics. Its non-minimal coupling might clarify flat galaxy rotation curves without presenting dark matter \cite{[c],[c1],[c2],[c3]}, frame-dragging, gyroscopic precession, and amplitude birefringence in the propagation and detection of gravitational waves.
	
	Among many modified gravity theories, the CS theory stands out as one of only a handful examples whose Dirichlet boundary problem has been well examined. Known answers for this hypothesis incorporate the Schwarzschild black hole and a slowly-rotating black hole. Loutrel and Tanaka \cite{[7a]} studied the late-time growth rates of the Wheeler-DeWitt patch for the Schwarzschild and slowly-rotating black holes using the Dirichlet boundary problem in the context of CS modified gravity.
	Utilizing holographic rule, 't Hooft \cite{[22m]} proposed an exceptionally straightforward and advantageous model to research the issues raised in DE, named holographic dark energy (HDE) model, and this model has recently gained much traction in the study of inflationary and late-time cosmic acceleration mechanisms \cite{no,nos1,nos2,noot, not}. This model is utilized in various situations with Hubble radius and cosmological conformal time of particle horizon \cite{[23a],[27]}. A fascinating Ricci holographic dark energy (RHDE) model characterized as $L= \mid R \mid^{-\frac{1}{2}}$, where $R$ is Ricci curvature scalar, was proposed by Gao et al \cite{[28]}.
	Silva and Santos \cite{[16a]} analyzed the RDE of the Friedmann-Lema\^{i}tre-Robertson-Walker (FLRW) universe and found it  to be similar to the generalized Chaplain gas (GCG) in the context of CS modified gravity. Jamil and Sarfraz \cite{[17]}  did similar work on the amended FLRW universe. Some of the present authors \cite{[17a]} also investigated cosmological parameters using the modified RHDE model and reconstructed different scalar field models such as
	quintessence, tachyon, k-essence and dilaton models in the context of CS modified gravity. Mathematically, it is observed that the EoS parameter predicted by the fact that dark energy is the dominant component of the universe is responsible for the accelerated expansion.
Jawad and Sohail \cite{[18]} considered modified QCD ghost dark energy models  to explore the elements of scalar field and possibilities of different
	scalar field models in the structure of dynamical CS modified gravity. Jamil and Sarfraz 
	\cite{[19]} considered the HDE model and found the accelerated expansion of the universe under specific limitations on the boundary $\alpha$. 
	Pasqua et al. \cite{[20]} investigated the HDE, modified holographic Ricci dark energy (MHRDE) and another model which is a combination of higher-order derivatives of the Hubble parameter in the framework of CS modified gravity.
A recently introduced Ricci-Gauss-Bonnet HDE model is studied in the context of CS modified gravity by Nasr \cite{[30a]}.
	
	Jawad et al. \cite{[31a]} found the accelerated expansion of the universe in the phantom as well as quintessence regions in the framework of DGP braneworld and dynamical CS modified gravity. They \cite{[31b]} also studied various modified entropies such as Bekenstein, logarithmic, power law correction and R\'{e}nyi along-with the investigations of first law of thermodynamics and generalized second law of thermodynamics on the apparent horizon. Porfirio and his team \cite{[32a]} discussed causality aspects of the dynamical CS modified gravity like relativistic fluid, cosmological constant, scalar and electromagnetic fields using G\"{o}del metric. In recent studies \cite{[7b]}-\cite{[7e]}, the HDE model has been considered broadly and analyzed as $\rho \propto \Lambda^{4}$ using the connections between IR, UV cutoff and the entropy such that $\Lambda^{3}L^{3}\leq S^{\frac{3}{4}}$. Working on the same lines, the relation of IR cut-offs along-with entropy gives rise to energy density of HDE model which depends on  the Bekenstein-Hawking entropy $S=\frac{A}{4G}$. The vacuum energy density is related with the UV cut-off Ricci scalar, particle horizon, Hubble horizon and event horizon, etc. i.e., the large-scale structure of the universe is related to the infrared (IR) cut-off. The HDE model perseveres through the choice of IR cut-off issue. A lot of literature is available on the investigations of different IR cut-offs in \cite{[8a]}-\cite{[8g]}. Some typical dark energy models, including $\Lambda$CDM, wCDM, CPL, and HDE models, are discussed by Zhang \cite{[8h]}.
	
	In the past few years, different entropies such as Tsallis \cite{[9a]}, R\'{e}nyi \cite{[9b]} and Sharma-Mittal \cite{[9c]} HDE models have been investigated for the cosmological and gravitational models. Chen \cite{[9d]} introduced recent developments on holographic entanglement entropy. Varying from regular HDE models with Bekenstein entropy, such models evolve to late-time accelerated universe. It is studied that the R\'{e}nyi HDE model exhibited a stable behavior in the case of a non-interacting universe \cite{[9b]}. Recently, Younus et. al  \cite{[9m]} studied Tsallis, R\'{e}nyi and Sharma-Mittal entropies and found quintessence-like nature of the universe in most of the cases. 
	Keeping in mind the above motivations, we investigated the R\'{e}nyi HDE using the FLRW metric in the context of CS modified gravity.
	
The article is organized as follows: in Sec. 2, the formalism of CS modified gravity theory and its field equations for FLRW metric are presented. In Sec. 3, we discuss R\'{e}nyi HDE model in the context of CS modified gravity theory in redshift space. Sec. 4 is devoted to exploring the stability of the system using the square of sound speed. Cosmic evolution is discussed in Section 5 and the last section is devoted to the discussions and conclusions.
	
	\section{The Chern-Simons formalism of modified  gravity}
	
	Among the myriad of possible modifications of General Relativity (GR)  is the CS modified gravity theory which is developed on the principle of leading-order gravitational parity violation. This modification is inspired by peculiarity cancellation in particle physics as well as string theory. The Einstein-Hilbert action is modified as
	\begin{equation}\label{csaction}
		S=S_{EH}+S_{CS}+S_{\Theta}+S_{mat},
	\end{equation}
	where Einstein-Hilbert term is denoted as
	\begin{equation}\
		S_{EH}= k\int_{v} d^{4}x \sqrt{-g}R,
	\end{equation}
	the CS term is represented as
	\begin{equation}
		\ S_{CS}= +\alpha\frac{1}{4}\int_{v} d^{4}x \sqrt{-g}\Theta  ^{\ast}RR,
	\end{equation}
	the scalar field is expressed
	\begin{equation}\
		S_{\Theta}= -\beta\frac{1}{2}\int_{v} d^{4}\times \sqrt{-g}[g^{ab}(\nabla_{a}\Theta)(\nabla_{b}\Theta)+2V(\Theta)],
	\end{equation}
	and the matter contribution to the action is given by
	\begin{equation}\
		S_{mat}= \int_{v} d^{4}x \sqrt{-g}\pounds_{mat},
	\end{equation}
	where $\pounds_{mat}$ represents some matter Lagrangian density, $k=\frac{1}{16\pi G}$, $g$ is  the determinant of the metric, $\nabla_{a}$ is the covariant derivative, R is the Ricci scalar and integrals represent the volume executed anywhere on the manifold $\nu$. The Pontryagin density $^{\ast}RR$is mathematically given as $^{\ast}RR=R\tilde{R}=^{\ast}R^{acd}_{b}R^{b}_{acd}$ and the dual Riemannian tensor is defined as
	$^{\ast}R^{a cd}_{b}=\frac{1}{2}\epsilon^{cdef}R^{a}_{bef},$
	where $\epsilon^{cdef}$ is the four dimensional Levi-Civita tensor. Formally, $^{\ast}RR \propto R\wedge R $, however, the curvature tensor is supposed to be
	the Riemannian tensor.
	
	The variation of action of Eq. \eqref{csaction} w.r.t. metric $g_{ab}$ and scalar field $\Theta$ gives  a set of field equations of CS modified gravity in the following forms:
	\ber
	&&G_{ab}+\alpha C_{ab}=-\frac{1}{2k}(T^{m}_{ab}+T^{\theta}_{ab})\;,\\
	&&\label{nbtheta}g^{ab}\nabla_{a}\nabla_{b}\Theta=-\frac{k \alpha}{4}~~^{\ast}RR,
	\eer
	$G_{ab}$ is Einstein tensor, $\alpha$ coupling constant, $C_{ab}$ is the Cotton tensor defined as
	\begin{equation}
		C^{ab}=-\frac{1}{2\sqrt{-g}}[\upsilon_{\sigma} \epsilon^{\sigma a\zeta\eta}\nabla_{\zeta}R^{b}_{\eta}+\frac{1}{2}\upsilon_{\sigma\tau}\epsilon^{\sigma b \zeta\eta}R^{\tau a}_{\zeta\eta}],
	\end{equation}
	where $\upsilon_{\sigma}=\nabla_{\sigma}\Theta$ and $\upsilon_{\sigma\tau}=\nabla_{\sigma}\nabla_{\tau}\Theta$. The energy-momentum tensor $T_{ab}$ comprises of matter and external field parts defined as:
	\ber\label{tm}
	&&T^{m}_{ab}=(\rho+p)U_{a}U_{b}-pg_{ab}\;,\\
	&&\label{ttheta}T^{\Theta}_{ab}=\eta(\partial_{a}\Theta)(\partial_{b}\Theta)-\frac{\eta}{2}g_{ab}(\partial^{\alpha}\Theta)(\partial_{\alpha}\Theta)\;.
	\eer
	Here p, $\rho$ and U are pressure, energy density and the four-vector velocity in co-moving coordinates of the spacetime respectively.\\
	Most of the study of the CS gravitational modification is in the non-dynamical framework. In this context, the scalar field is a non-dynamical term and it is supposed to be the prior prescriptive function of spacetime. These types of investigations introduce the evaluation of approximate solutions, exact solutions, cosmology, astrophysical tests and matter interactions. In non-dynamical CS modified gravity, there is a theoretical problem associated with Schwarzschild black hole perturbation theory, the occurrence of static and axisymmetric solutions and the uniqueness of solutions. However, there are numerous issues in non-dynamical theory such as:
	\begin{enumerate}[label=(\roman*)]
		\item  During the rotation of black hole singularities of curvature to be seen on the rotational axis; 
		\item Oscillation modes of the massive group of black holes are hidden;
		\item Commonly ghosts arise. 
	\end{enumerate}
	Therefore, the dynamical CS modified gravity including the kinetic term for the scalar field is prescribed by several authors to overcome the above mentioned issues and to preserve the self-stability of the theory. 
	
	The first two problems mentioned above do not arise in the dynamic theory and the last one does not occur in a particular condition. Consequently, the dynamical CS modified gravity has captured more interest in recent years.
	
	The FLRW model is considered the standard model of contemporary cosmology. Mathematically it can be written as
	\begin{eqnarray}\label{frw}
		ds^2=-dt^2+a^2(t)\left[\frac{dr^2}{1-\kappa r^2}+r^2(d\theta^2+{\sin^2}\theta d\phi^2)\right]\;,
	\end{eqnarray}
	where $a(t)$ is a scale factor, $t$ is the cosmic time parameter, $r$ represents the radial component and $\theta$ and $\phi$ are angular coordinates. $\kappa$ is known as the Gaussian curvature and the universe is considered open, closed or flat for $\kappa=-1,1,0$ respectively.
	
	The non-vanishing components of Einstein tensor for the metric given in Eq. \eqref{frw} are calculated as:
	\begin{eqnarray}
		G_{00}&=&\frac{3}{a^2}(\dot{a}^2+\kappa)\;,\nonumber\\
		G_{11}&=&-\left(\frac{2a\ddot{a}+\dot{a}^2+\kappa}{1-\kappa r^2}\right)\;,\nonumber\\
		G_{22}&=&-r^2(\kappa+\dot{a}^2+2a\ddot{a})\;,\nonumber\\
		G_{33}&=&-r^2\sin^2\theta(\kappa+\dot{a}^2+2a\ddot{a})\;.
	\end{eqnarray} 
	Using Eq. \eqref{ttheta}, the $00$-component of energy-momentum tensor of the external field is evaluated as
	\begin{eqnarray}
		\rho_{\Theta}=T^{\Theta}_{00} = \frac{\dot{\Theta}^2}{2}\;.
	\end{eqnarray}
	Fortunately, for the FLRW metric all the components of the Cotton tensor vanish identically:
	\begin{eqnarray}
		C_{ab}=0\;.
	\end{eqnarray}
	The term $^*RR$ is also zero for the  FLRW metric. It is a necessary condition for a metric to be a solution of Einstein field equations that the Pontryagin term $^*RR=0$. So, Eq. \eqref{nbtheta} reduces to
	\begin{eqnarray}\label{nbnbtheta}
		g^{ab}\nabla_{a}\nabla_{b}\Theta = g^{ab}[\partial_{a}\partial_{b}{\Theta }- \Gamma^{\rho}_{ab}{\partial_{\rho}\Theta}]=0\;.
	\end{eqnarray}
	The external field $\Theta$ is a function of spacetime coordinates in general, but for the sake of simplicity, it is assumed  to be a function of time parameter only and hence Eq. \eqref{nbnbtheta} gives the result in the following form:  
	\begin{eqnarray}\label{ddottheta}
		\ddot{\Theta} + 3 \frac{\dot{a}}{a} \dot{\Theta} = 0\;.
	\end{eqnarray}
	Using the method of separation of variables, we solve the differential Eq. \eqref{ddottheta} to obtain
	\begin{eqnarray}
		\dot{\Theta} = \rho_{\Theta_0}  a^{-3}\;,
	\end{eqnarray}
	where $\rho_{\Theta_0}$ is a constant of integration.

	\section{R\'{e}nyi Holographic Dark Energy Model}
	
	In modern cosmology, the origin and nature of dark energy is a mysterious issue. A lot of solutions for this problem are presented and the holographic dark energy (HDE) model is one of them. According to this model $\rho_{DE}=3d^2 M^2_p L^{-2}$,
where $d$ is a numerical factor, $M_p$ is the reduced Planck's mass and $L$ is the event horizon radius of the universe. It is worth noting here that there appears a strong correlation between quantum gravity and generalized entropy scenarios, and indeed, quantum aspects of universal gravitation may be considered as another motive to recognize entropies \cite{[80]}. The entropy of Tsallis is one of the importantentropy measures leading to acceptable results in different cosmological conditions \cite{[23s],[25s]},\cite{[81]}-\cite{[89]}. Tsallis entropy is usually defined as  \cite{[86]} for a discrete states system given by
	\begin{eqnarray}\label{abc1}
		S_{TE}=\frac{1}{1-N}\Sigma_{i=1}^{W}(P_{i}^N-P_{i}),
	\end{eqnarray}
	where $P_i$ is the ordinary probability of accessing state $i$ and $P$ is a real parameter that can be derived from nonextensive system features such as gravity's long-range nature \cite{[83],[86]}. In fact, nonextensivity is a more complicated concept than nonadditivity \cite{[87]}. For example, the well-known Bekenstein entropy is both nonadditive and nonextensive \cite{[84],[85]}. Recently, it has been proposed that the Bekenstein entropy  $S=\frac{A}{4}$ where $A=4\pi L^{2}$ and $L$ is the IR cut-off is actually a Tsallis entropy, which leads to the system's R\'{e}nyi entropy \cite{[21]} contents given by
	\begin{eqnarray}\label{abc9}
		S=\frac{1}{\delta}\log(\frac{\delta}{4}A+1)=\frac{1}{\delta}\log(\pi \delta L^{2}+1).
	\end{eqnarray}
	In this case, $\delta$ is a free parameter known in the literature as the real nonextensive parameter, which quantifies the degree of nonextensibility  \cite{[86],[87]}. Barboza et. al \cite{[90]} proposed that the $\delta$ parameter influences the Universe's energy balance. When $\delta<1$, the gravitational field is strong enough that we only need a small amount of DE and DM to construct the observable Universe. When $\delta>1$, on the other hand, the gravitational field is weak enough that we require a greater quantity of DE and DM. To summarise, this $\delta <1$ implies less DE and $\delta >1 $ implies more DE than in the standard Boltzmann-Gibbs scenario \cite{[90],[91]}.
In this model, the vacuum energy density plays the role of dark energy, that is, one has $\Lambda^4 \sim \rho_{\Lambda} \equiv \rho_{D}$. Using Eq. \eqref{abc9} and the assumptions $\rho_d \propto dS$ and $L = H^{-1}$ (i.e., Hubble horizon), the energy density for the  R\'{e}nyi \cite{[9b]} HDE model is defined as
	\begin{eqnarray}\label{rhor}
		\rho_{d}= \frac{3C^2H^2}{8\pi(1+\frac{\delta\pi}{H^2})}\;, 
	\end{eqnarray}
	where $H=\frac{\dot{a}}{a}$ is a Hubble parameter and $C$ is a constant.
	By law of conservation of energy-momentum, the matter density is given by $\rho_m= \rho_0a^{-3}$.
	For flat FLRW universe, CS field equations are tabled as:
	\begin{eqnarray}
		&&H^2=\frac{8\pi}{3}(\rho_{m}+\rho_{d}+\rho_\Theta),\\
		&&\label{h2p}H^2+\frac{2\dot{H}}{3}=-\frac{8\pi}{3}P_d\;.
	\end{eqnarray}
	Substituting the values of $\rho_{d}$, $\rho_{m}$, $\rho_{\Theta}$ in Eq.(19), one obtains the following expression:
	\begin{eqnarray}\label{hubblesq}
		H^2=\frac{8\pi}{3}\rho_0a^{-3}+\frac{C^2H^2}{1+\frac{\delta\pi}{H^2}}+\frac{4\pi}{3} {\rho_\Theta}_{0}a^{-6}.
	\end{eqnarray}
	Let us now define the normalized Hubble parameter $h(z)=\frac{H(z)}{H_0}$ and use $ z+1=a^{-1} $ in Eq. \eqref{hubblesq} to get the following equation:
	\begin{eqnarray}\label{hsq}
		h^2(z)=\Omega_m(z+1)^3+\left(\frac{C^2h^2(z)}{1+\frac{\delta\pi}{h^2(z)H^2_0}}\right)+\Omega_\Theta(z+1)^6\;.
	\end{eqnarray}
	Here, $\frac{8\pi\rho_0}{3H^2_0}=\Omega_m$ and $ \frac{{4\pi\rho_\Theta}_{0}}{3H^2_0}=\Omega_\Theta $, $z$ is the redshift parameter and $H_0$ is the value of Hubble parameter at $z=0$ respectively. 
	Using the initial conditions $z=0 $ and $h(z)=1 $, one can see that $ C^2=\{1-(\Omega_m+\Omega_\Theta)\}\left(1+\frac{\delta\pi}{H^2_0}\right)$.
	Using the value of $C^2$ in Eq. \eqref{rhor} the expression for R\'{e}nyi HDE model is given as
	\begin{eqnarray}
		\rho_{d}= \frac{3H^2}{8\pi\left(1+\frac{\delta\pi}{H^2}\right)}\{1-(\Omega_m+\Omega_\Theta)\}\left(1+\frac{\delta\pi}{H^2_0}\right)\;.
	\end{eqnarray}

	To investigate the stability of the R\'{e}nyi HDE model, one must study the evolution of square of sound speed defined as
	$c^2_s=\frac{dP_d}{d\rho_{d}}=\frac{\frac{d}{dH}(P_d)}{\frac{d}{dh}(\rho_{d})}$.
	Using  Eqs. \eqref{rhor} and \eqref{h2p}, it can be shown that the square speed of sound is positive, depicting the stability of the model but values of the parameter space for which $c^2_s>1$ might indicate breakdown of the causality in such instances.
	
	\section{Cosmic Evolution}
	
To understand the structure and dynamics of an astrophysical system, the EoS is usually necessary. By EoS of the universe, we mean the relation between the total energy density of cosmic matter and the total pressure. Another model called deceleration parameter is a dimensionless measure of the cosmic acceleration of the expansion of space studied in cosmology.	
	To study about the expansion rate of the universe, the value of $h(z)$ is calculated using Eq. \eqref{hsq}
	\begin{eqnarray}\label{sspeed}
		h^2(z)(1-C^2)+\frac{\delta\pi}{H^2_0}=\{(z+1)^3\Omega_m+(z+1)^6\Omega_\Theta\}\left(\frac{\frac{\delta\pi}{H^2_0}+h^2(z)}{h^2(z)}\right).
	\end{eqnarray}
	This equation  gives rise to two solutions and we will discuss them as two separate cases as follows.
	
	\subsection{Case 1}
	Taking into account the positive root of the quadratic equation in $h(z)$ in Eq. \eqref{sspeed}
	\begin{eqnarray}\label{h2sq}
		h(z)&=&\left[\frac{\Omega_m(z+1)^3+\Omega_\Theta(z+1)^6}{1-C^2}\right]^\frac{1}{2}\nonumber\\
		&\times& \left[\frac{\left(\frac{\delta\pi}{H^2_0}(1-C^2)\right)+\left(\Omega_m(z+1)^3+\Omega_\Theta(z+1)^6\right)}{\left(\frac{\delta\pi}{H^2_0}\right)+(\Omega_m(z+1)^3+\Omega_\Theta(z+1)^6)}\right]^\frac{1}{2}\;.
	\end{eqnarray}
	
	Differentiating Eq. \eqref{h2sq} with respect to $z$ and simplifying gives
	\begin{eqnarray}\label{dhz}
		\frac{d}{dz}h(z)&=&\frac{\left(\frac{\delta\pi}{H^2_0}\right)^2\{3\Omega_m(z+1)^2+6\Omega_\Theta(z+1)^5\}\sqrt{1-C^2}}{2\left(\frac{\delta\pi}{H^2_0}+\Omega_m.(z+1)^3+\Omega_\Theta.(z+1)^6\right)^\frac{3}{2}
		}\nonumber\\
		&+&\frac{\{\Omega_m(z+1)^3+\Omega_\Theta(z+1)^6\}\{3\Omega_m(z+1)^2+6\Omega_\Theta(z+1)^5\}\left(\frac{\delta\pi}{H^2_0}\right)}{\sqrt{1-C^2}{\left(\frac{\delta\pi}{H^2_0}+\Omega_m(z+1)^3+\Omega_\Theta(z+1)^6\right)^\frac{3}{2}
		}}\nonumber\\
		&+&\frac{\Omega_m(z+1)^3+\Omega_\Theta(z+1)^6}{{2\sqrt{1-C^2}\left(\frac{\delta\pi}{H^2_0}+\Omega_m(z+1)^3+\Omega_\Theta(z+1)^6\right)^\frac{3}{2}
		}}\;.
	\end{eqnarray}

	\subsubsection{Deceleration Parameter}
	
\begin{figure}[ht]
	\centering
	\includegraphics[width=3.5in]{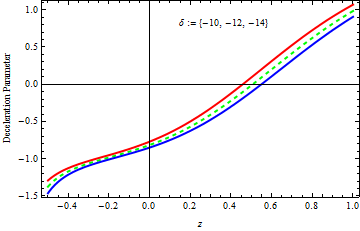}	
	\caption{Deceleration parameter $q $ versus redshift $z$ for Case 1.}
	\label{fq1}	 
\end{figure}

It is a dimensionless quantity in cosmology denoted by $q$ and mathematically defined as $q=-\frac{\ddot{a}a}{\dot{a}^2}$.
If $\ddot{a}>0$ and $q<0$, then the expansion rate of the universe will be accelerated.
The most familiar method of achieving such type of a scenario is to consider the
parametrization for the deceleration parameter as a function
of the scale factor $a$ or the redshift $z$ or the cosmic time $t$. The deceleration parameter in terms of Hubble scale parameter is evaluated as:
\begin{eqnarray}\label{decp}
	q=-1-\frac{\dot{H}}{H^2}.
\end{eqnarray}
Differentiating with respect to $z$, we obtain $\dot{H}=H_0\frac{d}{dz}h(z)$ and Eq. \eqref{decp} becomes
\begin{eqnarray}\label{dz}
	q=-1+\left(\frac{1+z}{h(z)}\right).\frac{d}{dz}(h(z)\;.
\end{eqnarray}
Substituting Eq. \eqref{h2sq} and Eq. \eqref{dhz} in Eq. \eqref{dz}, one obtains the following expression:
\begin{eqnarray}\label{q2}
	q&=&\frac{\Omega_m+4\Omega_\Theta(z+1)^3}{2\left[\Omega_m+\Omega_\Theta(z+1)^3\right]}\nonumber\\
	&+&
	\frac{(z+1)^3}{2}\left[\frac{\left(C^2\frac{\delta\pi}{H^2_0}\right)(3\Omega_m+6\Omega_\Theta(z+1)^3)}{\left(\frac{\delta\pi}{H^2_0}+\Omega_m(z+1)^3+\Omega_\Theta(z+1)^6\right)} \right]\nonumber\\
	&\times&\left[\frac{1}{\frac{\delta\pi}{H^2_0}+\Omega_m(z+1)^3+\Omega_\Theta(z+1)^6-C^2\frac{\delta\pi}{H^2_0}}\right]\;.
\end{eqnarray}
To investigate the deceleration parameter using R\'{e}nyi HDE in the context of CS modified gravity, we plotted a graph shown in Fig. \ref{fq1}. We opted the restrictions on parameters $\Omega_m=0.315$, $ \Omega_\Theta= 0.12$ , $ C=18 $, $ H_0=71~ km/s/Mpc$.
The graph illustrates the decelerated phase $q< 0$ for $-1<z<0.5$ and accelerated phase $q > 0$  for $0.5<z<1$ in Fig. \ref{fq1}. Also, it is observed that the behavior of the deceleration parameter is very similar in all three cases and our graphical representation advocates the transition from deceleration to accelerated phase which is fully consistent with the observational data.

	\subsubsection{Jerk Parameter}
	The jerk parameter is one of the higher-order expansion rate parameters frequently studied in cosmology and is defined as
\begin{equation}
	j=\frac{a^2}{\dot{a}^3}\frac{d^3a}{dt^3}=\frac{\ddot{H}}{H^3}-3q-2\;,
\end{equation}
which, in normalized parameters in redshift space, can be re-written as
\begin{equation}
	j=(1+z)\left[(1+z)\frac{h''}{h}+(1+z)\frac{h'^2}{h^2}+\frac{h'}{h}\right]-3q-2\;.
\end{equation}

\begin{figure}[ht]
		\centering
		\includegraphics[width=4in]{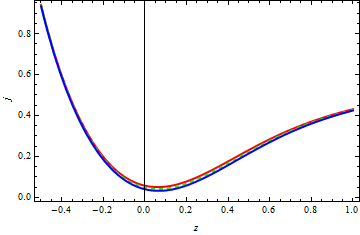}	
		\caption{jerk parameter $j$ versus redshift $z$ for Case 1.}
		\label{fq10}
	\end{figure}

To investigate the jerk parameter using R\'{e}nyi HDE in the context of CS modified gravity, we plotted a graph shown in Fig. \ref{fq10}. We opted the restrictions on parameters $\Omega_m=0.315$, $ \Omega_\Theta= 0.12$ , $ C=18 $, $ H_0=71~ km/s/Mpc$. It is observed that jerk parameter stays positive.
	
	\subsubsection{Equation of State}
	
\begin{figure}[ht]
		\centering
		\includegraphics[width=4in]{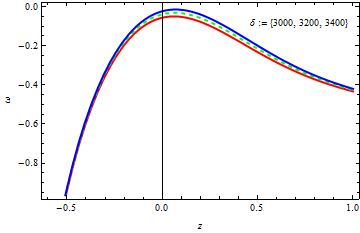}	
		\caption{The evolution of EoS $w $ versus redshift $ z$ for Case 1.}	 
		\label{feos1}
	\end{figure}

	The EoS represented by $\omega$ is a ratio of pressure $P$ and energy density $\rho$,  mathematically,
	$ \omega=\frac{P}{\rho} $. 
	One can obtain relation between $\omega$ and deceleration parameter $q$ using Eq. \eqref{h2p} and Eq. \eqref{decp} as
	\begin{eqnarray}\label{om}
		\omega=\frac{2}{3}\left(q-\frac{1}{2}\right)\;.
	\end{eqnarray}
	Making use of Eq. \eqref{q2} in Eq. \eqref{om},  $\omega$ is expressed as
	\begin{eqnarray}\label{Am2}
		\omega&=&\left(\frac{-z\Omega_m-(z+1)^4\Omega_\Theta+2(z+1)^3\Omega_\Theta}{\Omega_m(z+1)+\Omega_\Theta(z+1)^4}\right)+\frac{1+z}{2}\nonumber\\
		&\times&
		\left[\frac{\left(C^2\frac{\delta\pi}{H^2_0}\right)(3\Omega_m(z+1)^2+6\Omega_\Theta(z+1)^5)}{\left(\frac{\delta\pi}{H^2_0}+\Omega_m(z+1)^3+\Omega_\Theta(z+1)^6\right)} \right]\nonumber\\
		&\times&\left[\frac{1}{\frac{\delta\pi}{H^2_0}+\Omega_m(z+1)^3+\Omega_\Theta(z+1)^6-C^2\frac{\delta\pi}{H^2_0}}\right]\;.
	\end{eqnarray}
	From Eq. \eqref{Am2} it is obvious that the EoS $\omega$ is a function of $z$ along-with dependence on some cosmological parameters. To understand the cosmological evolution of the equation state, we plot the graph given in Fig. \ref{feos1}.
	
	\FloatBarrier
	The particular restrictions are imposed on the parameters such as $\Omega_m=0.315$, $ \Omega_\Theta= 0.25$, $ C=18 $, $ H_0=71~ km/s/Mpc$ and three different values of $ \delta $. In fact, EoS illustrates the era of dominance of the universe by certain components. For example, $\omega=0,\frac{1}{3}$ and $1$ predict that the universe is under dust, radiation and stiff fluid influence respectively. While $\omega=-\frac{1}{3}\;,-1$ and $\omega<-1$ stand for quintessence DE, $\Lambda$CDM and Phantom eras respectively. From the graph it is clear that the universe is under the influence of dark energy as the EoS predicted accelerated expansion phase.
	
\subsubsection{The $\omega-{\omega^{\prime}}$ Plane}
	
Caldwell and Linder \cite{[62]} introduced the $\omega-\omega^{\prime}$ plane to study the cosmic evolution of the quintessence dark energy model. They made a conclusion that the area contained by any dark energy model in this plane is divided into freezing ($\omega <0,~ \omega^{\prime}<0$) and thawing ($\omega<0,~ \omega^{\prime}>0$) regions. It is observed that the cosmic expansion of the universe is more accelerating in the freezing region as compared to thawing.
	
Differentiating Eq. \eqref{Am2} w.r.t. $z$, one gets
\begin{eqnarray}
{\omega^{\prime}}&=&\left[\frac{\{2(z+1)^2\}(\Omega_m+5\Omega_\Theta(z+1)^3).\frac{\delta\pi}{H^2_0}C^2}{ \frac{\delta\pi}{H^2_0}+\Omega_m(z+1)^3+\Omega_\Theta(z+1)^6(\frac{\delta\pi}{H^2_0}+\Omega_m(z+1)^3+\Omega_\Theta(z+1)^6)-C^2\frac{\delta\pi}{H^2_0}}\right]\nonumber\\
&-&
\frac{1}{(z+1)^3}\left[\frac{\{\Omega_m+2\Omega_\Theta(z+1)^3\}}{(\Omega_m+\Omega_\Theta(z+1)^3)^2}\right]+\frac{1}{1+z}\left[\frac{\Omega_m+2\Omega_\Theta(z+1)^3}{\Omega_m+\Omega_\Theta(z+1)^3}\right]\nonumber\\
&-&
\left[\frac{3(z+1)^5 (\Omega_m+2\Omega_\Theta(z+1)^3)^2.\frac{\delta\pi}{H^2_0}C^2}{\frac{\delta\pi}{H^2_0}+\Omega_m(z+1)^3+\Omega_\Theta(z+1)^6-C^2\frac{\delta\pi}{H^2_0}.(\frac{\delta\pi}{H^2_0}+\Omega_m(z+1)^3+\Omega_\Theta(z+1)^6)^2}\right]\nonumber\\
&-&
\left[\frac{\{3(z+1)^5\}(\Omega_m+2\Omega_\Theta(z+1)^3)^2.\frac{\delta\pi}{H^2_0}C^2}{ (\frac{\delta\pi}{H^2_0}+\Omega_m(z+1)^3+\Omega_\Theta(z+1)^6)-C^2\frac{\delta\pi}{H^2_0}(\frac{\delta\pi}{H^2_0}+\Omega_m(z+1)^3+\Omega_\Theta(z+1)^6) }\right]\nonumber\\
&+&
\frac{1}{(z+1)^7}\left[\frac{\{2\Omega_m+10\Omega_\Theta(z+1)^3\}\{\Omega_m+\Omega_\Theta(z+1)^3\}}{(\Omega_m+\Omega_\Theta(z+1)^3)^2}\right].
\end{eqnarray}
Taking into account the particular values of parameters $\Omega_m=0.315$, $\Omega_\Theta= 0.25$, $ C=20 $, $ H_0=67.4~ km/s/Mpc$ and three different values of $ \delta $, the following graph of $\omega^{\prime}$ is plotted.
	
	\begin{figure}[ht]
		\centering
		\includegraphics[width=4in]{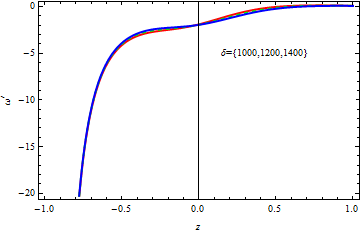}	
		\caption{ $\omega^{\prime}$ versus redshift $ z $ for Case 1.}	 
		\label{fomom1}
	\end{figure}
	\FloatBarrier
	The graphical behavior represented $ \omega <0,~ \omega^{\prime}<0$ indicated that the R\'{e}nyi HDE model is in freezing region and cosmic expansion is more accelerating in the context of CS modified gravity.
	
	\subsection{Case 2}
	The negative root of the Eq. \eqref{decp} is given by
	\begin{eqnarray}
		h(z)=\sqrt{\frac{\frac{\delta\pi}{H^2_0}}{1-C^2}\left[\frac{C^2\{\Omega_m(z+1)^3+\Omega_\Theta(z+1)^6\}}{\frac{\delta\pi}{H^2_0}+\Omega_m(z+1)^3+\Omega_\Theta(z+1)^6}\right]}\;.
	\end{eqnarray}
	Following similar procedures as in Case-1, we find the expression for the deceleration parameter $q$ given as
	\begin{eqnarray}\label{q3}
		q&=&-1+\left[\frac{(z+1)^3\left(3\Omega_m+6\Omega_\Theta(z+1)^3\right)}{\frac{\delta\pi}{H^2_0}+\Omega_m(z+1)^3+\Omega_\Theta(z+1)^6}\right]\nonumber\\
		&\times&
		\left[\frac{1}{(C^2-1)(\Omega_m(z+1)^3+\Omega_\Theta(z+1)^6)-\frac{\delta\pi}{H^2_0}}\right].
	\end{eqnarray}\\
	The graph of Eq. \eqref{q3} is plotted, with  the choice of parameter values $\Omega_m=0.315$, $ \Omega_\Theta= 0.25$, $ C=20 $, $ H_0=67.4 ~km/s/Mpc$ and $ \delta $ as shown below.
	\begin{figure}[ht]
		\centering
		\includegraphics[width=4in]{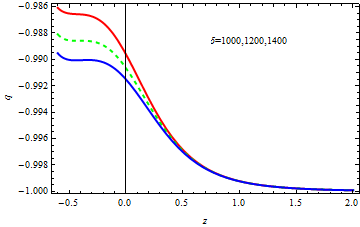}	
		\caption{The evolution of the deceleration parameter $q$ versus redshift $ z $ for Case 2.}
		\label{fq2}
		\label{fq}
	\end{figure}
	\FloatBarrier
	Fig. \eqref{fq} shows that the universe is in a decelerated phase of expansion as $q<0$ for each value of redshift parameter $z$ using R\'{e}nyi HDE in the context of CS modified gravity.
	
\begin{figure}[ht]
		\centering
		\includegraphics[width=4in]{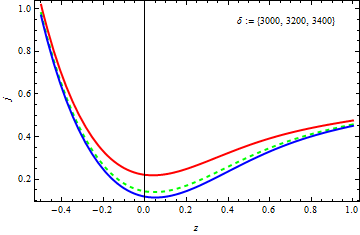}	
		\caption{jerk parameter $j$ versus redshift $z$ for Case 2.}
		\label{fq11}	 
	\end{figure}

	To investigate the jerk parameter case 2 we plotted a graph shown in Fig. \ref{fq11}. We opted the restrictions on parameters $\Omega_m=0.315$, $ \Omega_\Theta= 0.12$ , $ C=20 $, $ H_0=67.4~ km/s/Mpc$. It is observed that jerk parameter stays positive.
	\\
	Further, putting Eq. \eqref{q3} in Eq. \eqref{om}, the value of EoS $\omega$ is calculated as:
	\begin{eqnarray}\label{om2}
		\omega &=&-1+\left[\frac{2(z+1)^3\left(3\Omega_m+6\Omega_\Theta(z+1)^3\right)}{\frac{\delta\pi}{H^2_0}+\Omega_m(z+1)^3+\Omega_\Theta(z+1)^6}\right]\nonumber\\
		&\times&\left[\frac{1}{3(C^2-1)(\Omega_m(z+1)^3+\Omega_\Theta(z+1)^6)-\frac{\delta\pi}{H^2_0}}\right].
	\end{eqnarray}
	The graphical analysis of Eq. \eqref{om2} is plotted as follows.
	\begin{figure}[ht]
		\centering
		\includegraphics[width=4in]{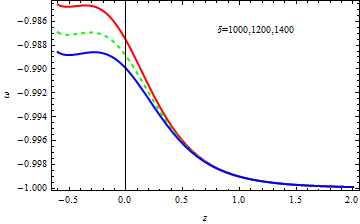}	
		\caption{The evolution of EoS versus redshift for Case 2.}
		\label{feos2}
	\end{figure}
	\FloatBarrier
	In Fig. \ref{feos2}, the parametric values are considered such that $\Omega_m=0.315$, $ \Omega_\Theta= 0.25$, $ C=20 $, $ H_0=67.4 ~km/s/Mpc$ and three different values of $ \delta $. The graph predicts that the universe is under the influence of $\Lambda$CDM.

Differentiating Eq. \eqref{om2} with respect to the redshift parameter $z$, gives the following expression:
	\begin{eqnarray}\label{omp}
		{\omega^{\prime}}&=&\frac{2(z+1)\left(6\Omega_m+30\Omega_\Theta(z+1)^4\right)+2\left(3\Omega_m(z+1)^2+6\Omega_\Theta(z+1)^5\right)}{3\left((C^2-1)(\Omega_m(z+1)^3+\Omega_\Theta(z+1)^6-\frac{\delta\pi}{H^2_0}\right)(\Omega_m(z+1)^3+\Omega_\Theta(z+1)^6+\frac{\delta\pi}{H^2_0})}\nonumber\\
		&-&\frac{2(C^2-1)(z+1)^5(3\Omega_m+6\Omega_\Theta(z+1)^3)^2}{ 3\left((C^2-1)(\Omega_m(z+1)^3+\Omega_\Theta(z+1)^6-\frac{\delta\pi}{H^2_0}\right)^2(\Omega_m(z+1)^3+\Omega_\Theta(z+1)^6+\frac{\delta\pi}{H^2_0})}\nonumber\\
		&-&\frac{2(z+1)^7(3\Omega_m+6\Omega_\Theta(z+1)^53)^2}{3\left((C^2-1)(\Omega_m(z+1)^3+\Omega_\Theta(z+1)^6-\frac{\delta\pi}{H^2_0}\right)^2(\Omega_m(z+1)^3+\Omega_\Theta(z+1)^6+\frac{\delta\pi}{H^2_0})^2}\;.\nonumber\\
	\end{eqnarray}
	The graph of Eq. \eqref{omp} is plotted under the restrictions on parameters $\Omega_m=0.315$, $ \Omega_\Theta= 0.25$, $ C=20 $, $ H_0=67.4 ~km/s/Mpc$ and $ \delta $ as shown below.
	\begin{figure}[ht]
		\centering
		\includegraphics[width=4in]{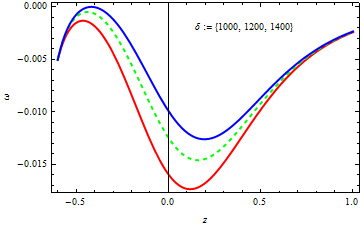}	
		\caption{$\omega^{\prime}$ versus redshift $ z $ for Case 2}
		\label{fomom2}
	\end{figure}
	\FloatBarrier
	In this case $ \omega <0,~ \omega^{\prime}<0$ indicates that the R\'{e}nyi HDE model is also in freezing region and cosmic expansion will be more accelerating in the context of CS modified gravity.
	
	\section{Conclusions}
	
In the present work, we studied the R\'{e}nyi HDE model in the context of CS modified gravity. We explored the EoS, deceleration parameter, cosmological plane and  square of sound speed for stability of the model in interacting scenarios. It is found that the value of $ c^2_s $ is positive for all real values of $ z $ which is a sufficient condition for system to be stable.
The CS field equations gave two different solutions and we discussed them separately. In the first case, the graph of the deceleration parameter illustrated the decelerated phase $q>0$ for $-1<z<0.6$ and accelerated phase $q < 0$ for $0.6<z<1$ in as depicted in Fig. \ref{fq1}. Also, it is observed that the behavior of deceleration parameter is very similar in all three values of $\delta$ of R\'{e}nyi HDE model and our graphical representation advocated the transition from deceleration to accelerated phase of the universe which is fully consistent with the observational data.
In fact, EoS illustrates the era of dominance of the universe by certain components. For example, $\omega=0,\frac{1}{3}$ and $1$ predicted that the universe is under dust, radiation and stiff fluid influence respectively. While $\omega=-\frac{1}{3},-1$ and $\omega<-1$ stand for quintessence DE, $\Lambda$CDM and Phantom eras respectively. From Fig. \ref{feos1} it is clear that the universe is under the influence of dark energy as the EoS predicted accelerated expansion phase. Caldwell and Linder \cite{[62]} introduced that the area contained by any dark energy model in the $\omega-\omega^{\prime}$ plane is divided into freezing and thawing regions. It is also observed that the cosmic expansion of the universe is more accelerating in the freezing region as compared to thawing.
The graphical behavior of Fig. \ref{fomom1} for $\omega <0,~ \omega^{\prime}<0$ indicated that the R\'{e}nyi HDE model is in freezing region, and cosmic expansion is more accelerating in the context of CS modified gravity.
In the second case, Fig. \ref{fq2} represented that the universe is in a decelerated phase of expansion as $q<0$ for each value of redshift parameter $z$ in the context of CS modified gravity. In Fig. \ref{feos2}, the graph of EoS predicted that the universe is under the influence of $\Lambda$CDM. Finally, the $\omega-\omega^{\prime}$ plane in Fig \ref{fomom2} indicated that the R\'{e}nyi HDE model is also in the freezing region and cosmic expansion will be more accelerating in the context of CS modified gravity. To fully understand the cosmological implications of this gravity theory in a broader context, we recognise the need to make a more comprehensive analysis - and constraining - of the parameter space allowed for the models we considered vis-\'a-vis existing and future data. This is left for future work.
	
	\section*{Acknowledgments}
	AA acknowledges that this work is base on the research supported in part by the NRF of South Africa with grant number 112131.

\end{document}